\documentclass{article} \usepackage{amsmath} \usepackage{hhline}
\usepackage{graphicx} \usepackage[T2A]{fontenc}
\usepackage[cp1251]{inputenc} \usepackage[russian]{babel}

\begin{document}          \parindent=13pt

\def\ds{\displaystyle}    \def\ss{\scriptstyle} \def\hh{\hskip 1pt}
\def\hs{\hskip 2pt}   \def\h{\hskip 0.2mm} \def\pr{\prime}
\newcommand{\blR}{\hbox{\ss\bf R}}
\newcommand{\fbr}{\hbox{\footnotesize\bf r}}
\newcommand{\fbk}{\hbox{\footnotesize\bf k}}
\newcommand{\Tr}{\mathop{\rm T\h r}\nolimits}
\newcommand{\fbR}{\hbox{\footnotesize\bf R}}

\setcounter{page}{1} \makeatletter\renewcommand{\@oddhead} { \hfil
---\hs\hs\hs\thepage\hs\hs\hs ---\hskip 2mm\hfil }
\makeatletter\renewcommand{\@oddfoot}{}

\hskip-13pt {\Large\bf The Fermi -- Dirac function and }

\vskip 2mm \hskip -13pt {\Large\bf the anisotropic distribution of
the interacting electrons } \vskip5mm

\hskip-13pt {\Large Boris V. Bondarev }\vskip 5mm

\hskip -13pt {\large\it Moscow Aviation Institute, Volokolamskoye
Shosse 4, 125871 Moscow, Russia} \vskip 1mm

\hskip -13pt {\large E-mail: bondarev.b@mail.ru} \vskip 10mm

\begin{quote} \hskip 7mm The distribution function for a system of
interacting electrons in metals is multivalent in a certain region
of wave vectors. One solution among many is isotropic. For other
solutions the distribution of electrons over the wave vectors is
anisotropic. In the simplest case, the anisotropy arises as a result
of the repulsion between electrons in states with the wave vectors
$\bf k$ and $-\hh\bf k$. \end{quote}

\vskip 0.2mm


\vskip 5mm \Large

\hskip -13pt {\large\bf 1.\hs Mean-field approximation }

\vskip 2mm The equilibrium  function $w_{\hh\fbk}$ of electron
distribution over the wave vectors in the mean field approximation
can be found from the equation [\hh 1 -- 3 ]:

$$ \ln\hs\frac{\hh 1-w_{\hh\fbk}\hh}{w_{\hh\fbk}}= \beta\hs
(\hh\overline\varepsilon_{\fbk}-\mu\hh )\hs , \eqno (1) $$ where
$w_{\hh\fbk}$ -- the probability of the state described by the wave
function $\psi_{\hh\fbk n}({\bf r})\hh\chi_\sigma(\xi)$ being
occupied by one of the electrons from the equilibrium system with
the wave vector $\bf k$; $\beta=1/k\hs T$, $T$ -- temperature,

$$ \overline\varepsilon_{\fbk}=\varepsilon_{\fbk} +
\sum\limits_{\fbk^\pr}\hs\varepsilon_{\fbk\fbk^\pr}\hs
w_{\hh\fbk^\pr} \eqno (2) $$ -- the average energy of an electron in
one of the states with the wave vector $\bf k$, $\varepsilon_{\fbk}$
-- the energy of electron without regard to its interaction with
other electrons, $\varepsilon_{\fbk\fbk^\pr}$ -- the interaction
energy of electrons with wave vectors $\bf k$ and $\bf k^\pr$, $\mu$
-- chemical potential. \par If the electrons do not interact with
each other, i.e. $\varepsilon_{\fbk\fbk^\pr}=0$, then from the
equation (1) we find the function of Fermi -- Dirac

$$ w_{\hh\fbk}=\frac{1}{\hs
1+e^{\hs\beta(\varepsilon_{\fbk}-\mu)}}\hs . $$

\newpage

\vskip 10mm\hskip -13pt {\bf 2. Repulsion of electrons with wave
vectors $\bf k$ and $-\hs\bf k$}

\vskip 2mm We assume the approximate formula for the interaction
energy

$$ \varepsilon_{\fbk\fbk^\pr}=I\hs\delta_{\hh\fbk\hh +\hh\fbk^\pr}
\hs , \eqno (3) $$ where $I$ -- a positive constant. According to
this approximate formula only those electrons interact - the wave
vectors of which are equal in absolute value and opposite in
direction: $\bf k^\pr=-\hh\bf k$. In this case $I$ is the
interaction energy of the electrons.

\par By the formula (2) the average energy of an electron will be
equal to

$$ \overline\varepsilon_{\fbk}=\varepsilon_{\fbk} + I\hs w_{\hh
-\fbk}\hs . \eqno (4) $$ According to this formula, the greater is
the probability $w_{\hh -\fbk}$ of the state with the wave vector
$-\hh\bf k$ being occupied, the larger is the energy of the electron
with the wave vector $\bf k$. Thus, the electron with the wave
vector $-\hh\bf k$ as if displaces the electron from the state with
the wave vector $\bf k$.

\par Put the expression (4) in the equation (1). We obtain the
equation (5)

$$ \ln\hs\frac{\hh 1-w_{\hh\fbk}\hh}{w_{\hh\fbk}}= \beta\hs
(\hh\varepsilon_{\fbk} + I\hs w_{\hh -\fbk}-\mu\hh )\hs . \eqno (5)
$$ Change the vector $\bf k$ to the vector $-\hh\bf k$ in the
equation (5) and obtain the equation (6)

$$ \ln\hs\frac{\hh 1-w_{\hh -\fbk}\hh}{w_{\hh -\fbk}}= \beta\hs
(\hh\varepsilon_{\fbk} + I\hs w_{\hh\fbk}-\mu\hh )\hs . \eqno (6) $$
The equations (5) and (6) form a system with two unknowns
$w_{\hh\fbk}$ and $w_{\hh -\fbk}$. It is easy to see that the
probability $w_{\hh\fbk}$ is a complex function of the vector $\bf
k$, in which the electron kinetic energy $\varepsilon_{\fbk}$ plays
the role of intermediate variable:

$$ w_{\hh\fbk}=w(\varepsilon_{\fbk})\hs . \eqno (7) $$ \par The
system of the equations (5) and (6) has two types of solutions. The
first are the isotropic distribution functions, which for all values
of the wave vector satisfy the condition

$$ w_{\hh -\fbk}=w_{\hh \fbk}\hs . \eqno (8) $$ The isotropic
distribution function is a solution of the equation

$$ \ln\hs\frac{\hh 1-w_{\hh\fbk}\hh}{w_{\hh\fbk}}= \beta\hs
(\hh\varepsilon_{\fbk} + I\hs w_{\hh\fbk}-\mu\hh )\hs . \eqno (9) $$

\par There are also anisotropic distribution functions for which the
condition (7) is not satisfied for some values of the wave vector:
$$ w_{\hh -\fbk}\neq w_{\hh \fbk}\hs . $$ And such anisotropy of the
electrons distribution over the Bloch states can take place even in
the absence of external fields.

\par Using the notation $$ w_{\hh -\fbk}=w_{\hh
1}(\varepsilon_{\hh\fbk})\hs , \hskip 7mm w_{\hh\fbk}=w_{\hh
2}(\varepsilon_{\hh\fbk})\hs , \eqno (10) $$ the equations (5) and
(6) can be written as:

$$ \left.\begin{array}{l} \ds\ln\hs\frac{\hh 1-w_{\hh 1}\hh}{w_{\hh
1}}= \frac{\hh 4\hh}{\tau}\hs(\hh\epsilon + w_{\hh 2}\hh )\hs ,
\medskip \\ \ds\ln\hs\frac{\hh 1-w_{\hh 2}\hh}{w_{\hh 2}}= \frac{\hh
4\hh}{\tau}\hs(\hh\epsilon + w_{\hh 1}\hh )\hs , \\
\end{array}\right\} \eqno (11) $$ where $$ \epsilon
=\frac{\hs\varepsilon -\mu}{I}\hs , \hskip 7mm \tau =\frac{\hs 4\hs
k\hs T \hs}{I}\hs . $$ In the equations (11) the unknowns are the
functions

$$ w_{\hh 1}=w_{\hh 1}(\varepsilon)\hskip 7mm\hbox{and}\hskip 7mm
w_{\hh 2}=w_{\hh 2}(\varepsilon)\hs . $$

\par When the electrons distribution over the wave vectors is
isotropic, we should put in the equations (11) $w_1=w_2=w_0$. It is
convenient to rewrite the obtained equation as:

$$ \epsilon=\frac{\tau}{\hh 4\hh}\hs\ln\hh \frac{\hs 1-w_0\hs}{\hs
w_0\hs}-w_0\hs . \eqno (12) $$ The equation (12) determines the
dependence $w_0=w_0(\epsilon)$, the graphs of which for different
values of temperature are shown in Fig.1 as monotonically decreasing
curves.

\par When the electrons distribution over the wave vectors is
anisotropic, in the equations (11) the probabilities $w_{\hh 1}$ and
$w_{\hh 2}$ should be considered as different functions of energy
$\epsilon$: $w_{\hh 1}=w_{\hh 1}(\epsilon)$, $w_{\hh 2}=w_{\hh
2}(\epsilon)$. In order to establish these relationships, we
introduce new variables $d$ and $s$ by means of relations

$$ w_2-w_1=d\hs , \hskip 7mm w_1+w_2=1+s\hs . \eqno (13) $$ Without
loss of the generality we can take the difference $d$ between two
values of the distribution functions $w_1$ and $w_2$ as the
non-negative value: $d\geq 0$, the greatest value of $d$ being equal
to one: $ d\in[\hh 0,\hh 1\hh]$. The parameter $s$ can take values
from $-1$ to 1: $s\in [\hh -1,\hh 1\hh]$. We solve the equation (13)
with respect to the probabilities $w_{\hh 1}$ and $w_{\hh 2}$:

$$ w_1=\frac{1}{\hh 2\hh}\hs (\hh 1+s-d\hh)\hs , \hskip 7mm
   w_2=\frac{1}{\hh 2\hh}\hs (\hh 1+s+d\hh)\hs . \eqno (14) $$

\par With the help of (14) we transform the system of the equations
(11). To do this  at first we subtract  one  equation from the other
and then add the equations. As a result, we obtain the following
system of the equations:

$$ \left.\begin{array}{l} \ds\frac{\hs (\hh 1+d\hh )^2-s^{\hh
2}\hh}{\hs (\hh 1-d\hh )^2-s^{\hh 2}\hh}= e^{\hh 4\hh d/\tau}\hs ,
\medskip \\ \ds \epsilon=\frac{\tau}{\hh 8\hh}\hs\ln\hh\frac{\hs
(\hh 1-s\hh )^2-d^{\hh 2}\hh} {\hs (\hh 1+s\hh )^2-d^{\hh
2}\hh}-\frac{1}{\hh 2\hh}\hs (\hh 1+s\hh)\hs . \\
\end{array}\right\} \eqno (15) $$ The first equation in this system
can be easily solved with respect to $s$:

$$ s(d)=\pm\hs\sqrt{\frac{\hh (\hh 1-d\hh)^{\hh 2}\hs e^{\hh 4\hh
d/\tau}- (\hh 1+d\hh)^{\hh 2}\hh}{e^{\hh 4\hh d/\tau}-1}}\hs . $$ By
virtue of relations (14) the probabilities of $w_{\hh 1}$ and
$w_{\hh 2}$ can also be viewed as the functions of the parameter
$d$: $w_{\hh 1}=w_{\hh 1}(d)$, $w_{\hh 2}=w_{\hh 2}(d)$. The second
equation of (15) allows us to express the energy $\epsilon$ of an
electron through the parameter $d$. With the help of the
dependencies obtained it is easy to construct the graphs of the
functions $w_{\hh 1}=w_{\hh 1}(\epsilon)$ and $w_{\hh 2}=w_{\hh
2}(\epsilon)$ for the different values of the temperature. Such
graphs are shown in the Fig. 2.

\par The character of the electrons distribution in Bloch states
depends on the relationship between the metal temperature $T$ and
the critical temperature

$$ T_c=\frac{I}{\hs 4\hs k\hh}\hs . \eqno (16) $$ At the
temperatures $T\geq T_c$ the distribution function $w=w(\epsilon)$
is single-valued and satisfies the condition (8) for all values of
the electron energies $\epsilon$.

\par At $T<T_c$ there is a range of the energy values
$(\epsilon_1,\hh\epsilon_2)$ at each point of which the function
$w=w(\epsilon)$ can take any of three values:
$w_1(\epsilon)<w_0(\epsilon)<w_2(\epsilon)$. Outside this interval,
the distribution function $w=w(\epsilon)$ takes only one value
$w_0(\epsilon)$. The function $w_{\hh\fbk}=w_0(\epsilon_{\hh\fbk})$
is a solution of the equation (12) and describes an isotropic
electron distribution over the wave vectors.

\par At $T<T_c$ in a narrow layer $S(\epsilon_1,\hh\epsilon_2)$
below the Fermi surface $\varepsilon_{\fbk}=\mu$  the anisotropic
electrons distribution over the wave vectors is possible. This
distribution is described by functions

$$ w_{\hh\fbk}=w_2(\epsilon_{\hh\fbk})\hs , \hskip 7mm
 w_{\hh -\fbk}=w_1(\epsilon_{\hh\fbk}) $$
at $\epsilon\in (\epsilon_1,\hh\epsilon_2)$.

\par At $T=0$ the isotropic distribution function has the form $$
w_{\hh\fbk}=\left\{\begin{array}{ccl} 1 & \hbox{at} &
\varepsilon_{\hh\fbk}\leq\mu - I\hs , \medskip \\
-\hs\ds\frac{1}{\hh I\hh}\hs (\hh\varepsilon_{\hh\fbk}-\mu\hh ) &
\hbox{at} & \mu -I<\varepsilon_{\hh\fbk}<\mu\hs , \hskip 3mm
\medskip \\ 0 & \hbox{at} & \varepsilon_{\hh\fbk}\geq\mu\hs ,
\end{array}\right. \eqno (17) $$ and the anisotropic distribution is
such that

$$ w_{\hh\fbk}=1 \hskip 4mm \hbox{at} \hskip 4mm
\varepsilon_{\hh\fbk}\leq\mu - I\hs , $$ $$ w_{\hh\fbk}=1\hs ,
\hskip 2mm w_{\hh -\fbk}=0 \hskip 4mm \hbox{or} \hskip 4mm
   w_{\hh\fbk}=0\hs , \hskip 2mm w_{\hh -\fbk}=1\hs
\hskip 4mm \hbox{at} \hskip 4mm \mu -I<\varepsilon_{\hh\fbk}<\mu\hs
, \eqno (18) $$ $$ w_{\hh\fbk}=0 \hskip 4mm \hbox{at} \hskip 4mm
\varepsilon_{\hh\fbk}\geq\mu\hs . $$ The formulae (18) show that
below the Fermi surface there is a layer $S$, defined by the
inequalities

$$ \mu -I<\varepsilon_{\hh\fbk}<\mu\hs , \eqno (19) $$ in which the
distribution of the electrons over the wave vectors is anisotropic,
i.e. of two states with wave vectors $\bf k$ and $-\hh\bf k$ in this
layer one is vacant and the other is necessarily occupied.

\par It can be shown that the electrons in an anisotropic
distribution have a lower energy than in the isotropic. Therefore,
in a layer of the electrons the anisotropic distribution is occupied
and the isotropic one is empty.

\par Calculation show that anisotropic solutions have lower energy
than the isotropic. Accordingly the real solutions are shown in Fig.
3 -- 5.

\vskip 4mm\hskip -13pt {\bf 3. Anisotropy parameter } \vskip 2mm

\par The difference $d=w_{\hh 2}-w_{\hh 1}$ between the two values
of the anisotropic distribution function takes its maximum value
$d_{\hh max}$ at $\epsilon=0.5$. At the same time $w_0=0.5$ and
$s=0$. The  dependence $d_{\hh max}$ on the temperature can be
found from the first equation of the system (15), putting in it
$s=0$:

$$ \frac{\hs 2\hs d_{\hh max}\hs}{\tau} =\ln\hh\frac{\hs 1+d_{\hh
max}\hh} {\hs 1-d_{\hh max}\hh}\hs . \eqno (20) $$ The graph of this
function is shown in the Fig. 6.

\vskip 4mm\hskip -13pt {\bf 4. The average energy of one electron }
\vskip 2mm

\par The dependence of the average energy
$\overline\varepsilon_{\fbk}$ of one electron on its kinetic energy
$\varepsilon_{\fbk}$ is given by (4). As it can be seen from this
formula, the energy of the electron with wave vector $\bf k$ depends
on whether a state with wave vector $-\hh\bf k$ is vacant or
occupied. The average energy of an electron $\overline\varepsilon$
can be represented as a function of its kinetic energy $\varepsilon$
as follows:

$$ \overline\varepsilon(\varepsilon)=\varepsilon + I\hs
w_1(\varepsilon)\hs . \eqno (21) $$ The graphs of this function for
different values of the temperature are shown in the Fig. 7.

\vskip 4mm\hskip -13pt {\bf 5.\hs Conclusions } \vskip 2mm

\par In this work we show the existence of the anisotropy in the
distribution of the interacting electrons over the wave vectors. In
the simplest case, such anisotropy arises as a result of the
repulsion between electrons in the states with the wave vectors $\bf
k$ and $-\hh\bf k$.

\par It can be shown that the anisotropy arises also in the other
dependencies of the interaction energy $\varepsilon_{\fbk\fbk^\pr}$
of the electrons. \par The following paper should be devoted to the
physical phenomena that are the consequences of the anisotropy of
the distribution functions.

\vskip 4mm\hskip -13pt {\bf References } \vskip 2mm

\hskip -13pt [1] J.V. Gibbs, The basis principles of statistical
mechanics (Ogiz, Moscow, 1946).

\vskip 0mm\hskip -13pt [2] L.D. Landau, E.M. Lifshitz, Statistical
physics (Nauka, Moscow, 1964).

\vskip 0mm\hskip -13pt [3] B.V. Bondarev, Vectnik MAI 3 (1996) 56.

\newpage

\parindent=13pt \def\ds{\displaystyle} \def\ss{\scriptsize}
\def\hh{\hskip 1pt}       \def\hs{\hskip 2pt} \def\h{\hskip 0.2mm}
\def\pr{\prime}

\setcounter{page}{7} \makeatletter\renewcommand{\@oddhead} { \hfil
---\hs\hs\hs\thepage\hs\hs\hs ---\hskip 2mm\hfil }
\makeatletter\renewcommand{\@oddfoot}{} \vskip 10mm

\hskip -13pt {\Large\bf The Fermi -- Dirac function and }

\vskip 2mm \hskip -13pt {\Large\bf the anisotropic distribution of
the interacting electrons }\vskip 5 mm

\hskip -13pt {\large Boris V. Bondarev } \unitlength=1mm

\vskip 20mm

\centerline{\unitlength=1.2mm\begin{picture}(87,57) \put(30,41){\it
1}\put(47,17){\it 1} \put(13,44){\it 2}\put(65,12.4){\it 2}
\put(4,42){\it 3} \put(75,14){\it 3}
\put(0,10){\vector(1,0){87}}\put(79,5){$\varepsilon-\mu$}
\multiput(20,10.5)(0,2){20}{\line(0,1){1.4}}
\multiput(20,50)(2,0){20}{\line(1,0){1.4}}
\put(20,10){\line(0,-1){1}}\put(17,5){$-I$}
\put(40,30){\circle*{0.7}}
\multiput(40,10.5)(0,2){10}{\line(0,1){1.4}}
\put(40,10){\line(0,-1){1}}\put(37,4.5){$-\frac12\hh I$}
\put(60,10){\line(0,-1){1}}\put(59.2,5){0}
\put(60,10){\vector(0,1){50}}\put(62,58.5){$w$} 
\put(0,50){\line(1,0){21}}\put(62,49){1,0}
\multiput(59.5,30)(-2,0){10}{\line(-1,0){1.4}}
\put(60,30){\line(1,0){1}}\put(62,29){0,5}
\put(60,10){\unitlength=1.2mm\special{em:linewidth
0.3pt}\put(-40,40) {\special{em:moveto}}
\put(0,0){\special{em:lineto}} }
\put(60,10){\unitlength=1.2mm\special{em:linewidth 0.3pt}
\put(11.62,0.3333)  {\special{em:moveto}} \put(9.528,0.6667)
{\special{em:lineto}} \put(8.160,1)  {\special{em:lineto}}
\put(5.360,2)  {\special{em:lineto}} \put(3.280,3)
{\special{em:lineto}} \put(1.492,4)  {\special{em:lineto}}
\put(-0.136,5)  {\special{em:lineto}} \put(-1.664,6)
{\special{em:lineto}} \put(-3.122,7)  {\special{em:lineto}}
\put(-4.536,8)  {\special{em:lineto}} \put(-5.908,9)
{\special{em:lineto}} \put(-7.252,10)  {\special{em:lineto}}
\put(-8.576,11)  {\special{em:lineto}} \put(-9.884,12)
{\special{em:lineto}} \put(-11.17,13)  {\special{em:lineto}}
\put(-12.45,14)  {\special{em:lineto}} \put(-13.72,15)
{\special{em:lineto}} \put(-14.99,16)  {\special{em:lineto}}
\put(-16.24,17)  {\special{em:lineto}} \put(-17.50,18)
{\special{em:lineto}} \put(-18.75,19)  {\special{em:lineto}}
\put(-20.00,20)  {\special{em:lineto}} \put(-21.25,21)
{\special{em:lineto}} \put(-22.50,22)  {\special{em:lineto}}
\put(-23.76,23)  {\special{em:lineto}} \put(-25.01,24)
{\special{em:lineto}} \put(-26.28,25)  {\special{em:lineto}}
\put(-27.55,26)  {\special{em:lineto}} \put(-28.83,27)
{\special{em:lineto}} \put(-30.12,28)  {\special{em:lineto}}
\put(-31.42,29)  {\special{em:lineto}} \put(-32.75,30)
{\special{em:lineto}} \put(-34.09,31)  {\special{em:lineto}}
\put(-35.46,32)  {\special{em:lineto}} \put(-36.88,33)
{\special{em:lineto}} \put(-38.34,34)  {\special{em:lineto}}
\put(-39.86,35)  {\special{em:lineto}} \put(-41.48,36)
{\special{em:lineto}} \put(-43.28,37)  {\special{em:lineto}}
\put(-45.36,38)  {\special{em:lineto}} \put(-48.16,39)
{\special{em:lineto}} \put(-49.52,39.33)  {\special{em:lineto}}
\put(-51.60,39.67)  {\special{em:lineto}} }
\put(60,10){\unitlength=1.2mm\special{em:linewidth 0.3pt}
\put(27.00,1.2)  {\special{em:moveto}}
\put(25.99,1.3){\special{em:lineto}} \put(21.55,2)
{\special{em:lineto}} \put(17.10,3)  {\special{em:lineto}}
\put(13.58,4)  {\special{em:lineto}} \put(10.57,5)
{\special{em:lineto}} \put(7.880,6)  {\special{em:lineto}}
\put(5.408,7)  {\special{em:lineto}} \put(3.088,8)
{\special{em:lineto}} \put(0.8960,9)  {\special{em:lineto}}
\put(-1.208,10)  {\special{em:lineto}} \put(-3.244,11)
{\special{em:lineto}} \put(-5.224,12)  {\special{em:lineto}}
\put(-7.152,13)  {\special{em:lineto}} \put(-9.048,14)
{\special{em:lineto}} \put(-10.91,15)  {\special{em:lineto}}
\put(-12.76,16)  {\special{em:lineto}} \put(-14.58,17)
{\special{em:lineto}} \put(-16.40,18)  {\special{em:lineto}}
\put(-18.20,19)  {\special{em:lineto}} \put(-20.00,20)
{\special{em:lineto}} \put(-21.80,21)  {\special{em:lineto}}
\put(-23.60,22)  {\special{em:lineto}} \put(-25.42,23)
{\special{em:lineto}} \put(-27.24,24)  {\special{em:lineto}}
\put(-29.09,25)  {\special{em:lineto}} \put(-30.95,26)
{\special{em:lineto}} \put(-32.85,27)  {\special{em:lineto}}
\put(-34.78,28)  {\special{em:lineto}} \put(-36.76,29)
{\special{em:lineto}} \put(-38.79,30)  {\special{em:lineto}}
\put(-40.88,31)  {\special{em:lineto}} \put(-43.08,32)
{\special{em:lineto}} \put(-45.40,33)  {\special{em:lineto}}
\put(-47.88,34)  {\special{em:lineto}} \put(-50.56,35)
{\special{em:lineto}} \put(-53.56,36)  {\special{em:lineto}}
\put(-57.08,37)  {\special{em:lineto}} \put(-60.00,37.7)
{\special{em:lineto}} } \end{picture}}

\vskip -5mm Fig. 1.

\vskip 15mm \centerline{\unitlength=1.2mm\begin{picture}(87,57)
\put(21,11){\it 1}\put(25,16){\it 2}\put(30.5,23.2){\it
3}\put(34,28){\it 4} \put(54,51){\it 1}\put(48.5,44.5){\it 2}
\put(0,10){\vector(1,0){87}}\put(79,5){$\varepsilon-\mu$}
\multiput(20,10.5)(0,2){20}{\line(0,1){1.4}}
\put(20,10){\line(0,-1){1}}\put(17,5){$-I$}
\put(40,30){\circle*{0.7}}\multiput(40,10.5)(0,2){10}{\line(0,1){1.4}}
\put(40,10){\line(0,-1){1}}\put(37,4.5){$-\frac12\hh I$}
\put(60,10){\line(0,-1){1}}\put(59.2,5){0}
\put(60,10){\vector(0,1){50}}\put(62,58.5){$w$} 
\put(20,50){\unitlength=1.2mm\line(1,0){41}}\put(62,49){1,0}
\put(20,50.2){\unitlength=1.2mm\line(1,0){40}}
\put(20,9.8){\unitlength=1.2mm\line(1,0){40}}
\multiput(20,50)(-2,0){10}{\line(-1,0){1.4}}
\multiput(59.5,30)(-2,0){10}{\line(-1,0){1.4}}
\put(60,30){\line(1,0){1}}\put(62,29){0,5}
\put(60,10){\unitlength=1.2mm\special{em:linewidth 0.3pt}
\put(-43.84,36.59)  {\special{em:moveto}} \put(-43.76,35.77)
{\special{em:lineto}} \put(-43.60,34.84)  {\special{em:lineto}}
\put(-43.40,33.82)  {\special{em:lineto}} \put(-43.16,32.73)
{\special{em:lineto}} \put(-42.88,31.57)  {\special{em:lineto}}
\put(-40.96,23.96)  {\special{em:lineto}} \put(-40.64,22.64)
{\special{em:lineto}} \put(-40.32,21.32)  {\special{em:lineto}}
\put(-39.98,19.99)  {\special{em:lineto}} \put(-39.66,18.66)
{\special{em:lineto}} \put(-39.32,17.33)  {\special{em:lineto}}
\put(-38.98,16.00)  {\special{em:lineto}} \put(-38.64,14.66)
{\special{em:lineto}} \put(-38.26,13.33)  {\special{em:lineto}}
\put(-37.88,12.00)  {\special{em:lineto}} \put(-37.47,10.66)
{\special{em:lineto}} \put(-37.02,9.336)  {\special{em:lineto}}
\put(-36.54,8.000)  {\special{em:lineto}} \put(-35.98,6.668)
{\special{em:lineto}} \put(-35.32,5.336)  {\special{em:lineto}}
\put(-34.51,4.000)  {\special{em:lineto}} \put(-33.41,2.664)
{\special{em:lineto}} \put(-32.06,1.600)  {\special{em:lineto}}
\put(-31.57,1.336)  {\special{em:lineto}} \put(-30.27,0.800)
{\special{em:lineto}} \put(-29.20,0.536)  {\special{em:lineto}}
\put(-27.56,0.264)  {\special{em:lineto}} \put(-1.100,0.004)
{\special{em:lineto}} \put(-0.344,0.004)  {\special{em:lineto}}
\put(0.3148,0.020)  {\special{em:lineto}} \put(0.9626,0.044)
{\special{em:lineto}} \put(1.6024,0.096)  {\special{em:lineto}}
\put(2.2572,0.208)  {\special{em:lineto}} \put(2.8704,0.428)
{\special{em:lineto}} \put(3.4012,0.845)  {\special{em:lineto}}
\put(3.7720,1.567)  {\special{em:lineto}} \put(3.8710,3.401)
{\special{em:lineto}} \put(3.7714,4.232)  {\special{em:lineto}}
\put(3.6118,5.162)  {\special{em:lineto}} \put(3.4014,6.181)
{\special{em:lineto}} \put(3.1508,7.275)  {\special{em:lineto}}
\put(2.8701,8.431)  {\special{em:lineto}} \put(2.5689,9.634)
{\special{em:lineto}} \put(2.2545,10.88)  {\special{em:lineto}}
\put(1.9330,12.14)  {\special{em:lineto}} \put(1.6077,13.43)
{\special{em:lineto}} \put(1.2823,14.73)  {\special{em:lineto}}
\put(0.9579,16.04)  {\special{em:lineto}} \put(0.6335,17.36)
{\special{em:lineto}} \put(0.3093,18.69)  {\special{em:lineto}}
\put(-0.014,20.01)  {\special{em:lineto}} \put(-0.344,21.34)
{\special{em:lineto}} \put(-0.669,22.67)  {\special{em:lineto}}
\put(-1.016,24.00)  {\special{em:lineto}} \put(-1.401,25.34)
{\special{em:lineto}} \put(-1.729,26.67)  {\special{em:lineto}}
\put(-2.127,28.00)  {\special{em:lineto}} \put(-2.523,29.33)
{\special{em:lineto}} \put(-2.899,30.67)  {\special{em:lineto}}
\put(-3.456,32.00)  {\special{em:lineto}} \put(-3.993,33.33)
{\special{em:lineto}} \put(-4.607,34.67)  {\special{em:lineto}}
\put(-5.484,36.00)  {\special{em:lineto}} \put(-6.555,37.33)
{\special{em:lineto}} \put(-7.088,37.87)  {\special{em:lineto}}
\put(-8.390,38.67)  {\special{em:lineto}} \put(-9.074,38.93)
{\special{em:lineto}} \put(-10.68,39.47)  {\special{em:lineto}}
\put(-12.56,39.73)  {\special{em:lineto}} \put(-38.98,40.00)
{\special{em:lineto}} \put(-39.66,40.00)  {\special{em:lineto}}
\put(-40.32,39.98)  {\special{em:lineto}} \put(-40.96,39.96)
{\special{em:lineto}} \put(-41.60,39.90)  {\special{em:lineto}}
\put(-42.24,39.79)  {\special{em:lineto}} \put(-42.88,39.57)
{\special{em:lineto}} \put(-43.40,39.16)  {\special{em:lineto}}
\put(-43.76,38.43)  {\special{em:lineto}} \put(-43.84,36.59)
{\special{em:lineto}} }
\put(60,10){\unitlength=1.2mm\special{em:linewidth 0.3pt}
\put(-36.61,28.25)  {\special{em:moveto}} \put(-36.56,27.54)
{\special{em:lineto}} \put(-36.49,26.81)  {\special{em:lineto}}
\put(-36.30,26.00)  {\special{em:lineto}} \put(-36.16,25.22)
{\special{em:lineto}} \put(-35.90,24.36)  {\special{em:lineto}}
\put(-35.60,23.46)  {\special{em:lineto}} \put(-35.30,22.58)
{\special{em:lineto}} \put(-34.92,21.62)  {\special{em:lineto}}
\put(-34.49,20.64)  {\special{em:lineto}} \put(-34.04,19.66)
{\special{em:lineto}} \put(-33.52,18.62)  {\special{em:lineto}}
\put(-32.93,17.55)  {\special{em:lineto}} \put(-32.27,16.47)
{\special{em:lineto}} \put(-31.54,15.34)  {\special{em:lineto}}
\put(-30.70,14.16)  {\special{em:lineto}} \put(-29.75,12.97)
{\special{em:lineto}} \put(-28.64,11.70)  {\special{em:lineto}}
\put(-27.28,10.36)  {\special{em:lineto}} \put(-25.57,8.904)
{\special{em:lineto}} \put(-23.82,7.696)  {\special{em:lineto}}
\put(-22.74,7.064)  {\special{em:lineto}} \put(-20.56,6.016)
{\special{em:lineto}} \put(-17.259,4.936)  {\special{em:lineto}}
\put(-13.507,4.348)  {\special{em:lineto}} \put(-11.362,4.300)
{\special{em:lineto}} \put(-9.7588,4.428)  {\special{em:lineto}}
\put(-8.4684,4.664)  {\special{em:lineto}} \put(-7.3924,4.988)
{\special{em:lineto}} \put(-6.4840,5.376)  {\special{em:lineto}}
\put(-5.7180,5.848)  {\special{em:lineto}} \put(-5.0736,6.376)
{\special{em:lineto}} \put(-4.5384,6.984)  {\special{em:lineto}}
\put(-4.1092,7.640)  {\special{em:lineto}} \put(-3.7788,8.388)
{\special{em:lineto}} \put(-3.5472,9.188)  {\special{em:lineto}}
\put(-3.4080,10.12)  {\special{em:lineto}} \put(-3.379, 11.74)
{\special{em:lineto}} \put(-3.4399,12.46)  {\special{em:lineto}}
\put(-3.5434,13.20)  {\special{em:lineto}} \put(-3.6889,13.99)
{\special{em:lineto}} \put(-3.8770,14.80)  {\special{em:lineto}}
\put(-4.1085,15.65)  {\special{em:lineto}} \put(-4.3842,16.53)
{\special{em:lineto}} \put(-4.7046,17.43)  {\special{em:lineto}}
\put(-5.0727,18.38)  {\special{em:lineto}} \put(-5.4901,19.35)
{\special{em:lineto}} \put(-5.9590,20.35)  {\special{em:lineto}}
\put(-6.4850,21.38)  {\special{em:lineto}} \put(-7.0731,22.44)
{\special{em:lineto}} \put(-7.7300,23.54)  {\special{em:lineto}}
\put(-8.4674,24.67)  {\special{em:lineto}} \put(-9.3005,25.83)
{\special{em:lineto}} \put(-10.252,27.04)  {\special{em:lineto}}
\put(-11.364,28.30)  {\special{em:lineto}} \put(-12.705,29.63)
{\special{em:lineto}} \put(-14.444,31.10)  {\special{em:lineto}}
\put(-15.347,31.76)  {\special{em:lineto}} \put(-16.176,32.30)
{\special{em:lineto}} \put(-17.259,32.94)  {\special{em:lineto}}
\put(-19.453,33.99)  {\special{em:lineto}} \put(-22.74,35.06)
{\special{em:lineto}} \put(-26.49,35.65)  {\special{em:lineto}}
\put(-28.64,35.70)  {\special{em:lineto}} \put(-30.24,35.57)
{\special{em:lineto}} \put(-31.54,35.34)  {\special{em:lineto}}
\put(-32.61,35.01)  {\special{em:lineto}} \put(-33.52,34.62)
{\special{em:lineto}} \put(-34.28,34.15)  {\special{em:lineto}}
\put(-34.92,33.62)  {\special{em:lineto}} \put(-35.45,33.02)
{\special{em:lineto}} \put(-35.90,32.36)  {\special{em:lineto}}
\put(-36.23,31.61)  {\special{em:lineto}} \put(-36.49,30.81)
{\special{em:lineto}} \put(-36.55,29.88)  {\special{em:lineto}}
\put(-36.61,28.25)  {\special{em:lineto}} }
\put(60,10){\unitlength=1.2mm\special{em:linewidth 0.3pt}
\put(-21.357,28.06)  {\special{em:moveto}} \put(-23.367,28.62)
{\special{em:lineto}} \put(-24.498,28.62)  {\special{em:lineto}}
\put(-25.346,28.60)  {\special{em:lineto}} \put(-26.022,28.47)
{\special{em:lineto}} \put(-26.574,28.24)  {\special{em:lineto}}
\put(-27.053,27.95)  {\special{em:lineto}} \put(-27.446,27.69)
{\special{em:lineto}} \put(-27.779,27.38)  {\special{em:lineto}}
\put(-28.059,27.04)  {\special{em:lineto}} \put(-28.299,26.66)
{\special{em:lineto}} \put(-28.460,26.28)  {\special{em:lineto}}
\put(-28.609,25.78)  {\special{em:lineto}} \put(-28.722,25.34)
{\special{em:lineto}} \put(-28.756,24.74)  {\special{em:lineto}}
\put(-28.756,23.74)  {\special{em:lineto}} \put(-28.722,23.34)
{\special{em:lineto}} \put(-28.609,22.78)  {\special{em:lineto}}
\put(-28.460,22.28)  {\special{em:lineto}} \put(-28.299,21.66)
{\special{em:lineto}} \put(-28.059,21.04)  {\special{em:lineto}}
\put(-27.779,20.38)  {\special{em:lineto}} \put(-27.446,19.69)
{\special{em:lineto}} \put(-27.053,18.95)  {\special{em:lineto}}
\put(-26.574,18.24)  {\special{em:lineto}} \put(-26.022,17.47)
{\special{em:lineto}} \put(-25.346,16.60)  {\special{em:lineto}}
\put(-24.498,15.62)  {\special{em:lineto}} \put(-23.367,14.62)
{\special{em:lineto}} \put(-21.357,13.06)  {\special{em:lineto}}
\put(-18.642,11.94)  {\special{em:lineto}} \put(-16.633,11.38)
{\special{em:lineto}} \put(-15.502,11.38)  {\special{em:lineto}}
\put(-14.653,11.40)  {\special{em:lineto}} \put(-13.979,11.54)
{\special{em:lineto}} \put(-13.427,11.76)  {\special{em:lineto}}
\put(-12.947,12.05)  {\special{em:lineto}} \put(-12.555,12.31)
{\special{em:lineto}} \put(-12.222,12.62)  {\special{em:lineto}}
\put(-11.941,12.96)  {\special{em:lineto}} \put(-11.700,13.34)
{\special{em:lineto}} \put(-11.541,13.72)  {\special{em:lineto}}
\put(-11.392,14.22)  {\special{em:lineto}} \put(-11.277,14.66)
{\special{em:lineto}} \put(-11.244,15.26)  {\special{em:lineto}}
\put(-11.244,16.26)  {\special{em:lineto}} \put(-11.277,16.66)
{\special{em:lineto}} \put(-11.392,17.22)  {\special{em:lineto}}
\put(-11.541,17.72)  {\special{em:lineto}} \put(-11.700,18.34)
{\special{em:lineto}} \put(-11.941,18.94)  {\special{em:lineto}}
\put(-12.222,19.62)  {\special{em:lineto}} \put(-12.555,20.31)
{\special{em:lineto}} \put(-12.947,21.05)  {\special{em:lineto}}
\put(-13.427,21.76)  {\special{em:lineto}} \put(-13.979,22.54)
{\special{em:lineto}} \put(-14.653,23.40)  {\special{em:lineto}}
\put(-15.502,24.38)  {\special{em:lineto}} \put(-16.633,25.38)
{\special{em:lineto}} \put(-18.642,26.94)  {\special{em:lineto}}
\put(-21.357,28.06)  {\special{em:lineto}} } \end{picture}}

\vskip -5mm Fig. 2.

\vskip 40mm

\vskip 15mm \centerline{\unitlength=1.2mm\begin{picture}(87,77)
\put(0,10){\vector(1,0){87}}\put(79,5){$\varepsilon-\mu$}
\multiput(20,10.5)(0,2){20} {\line(0,1){1.4}}
\put(20,10){\line(0,-1){1}}\put(17,5){$-I$}
\put(40,10){\line(0,-1){1}}
\put(60,10){\line(0,-1){1}}\put(59.2,5){0}
\put(60,10){\vector(0,1){50}}\put(62,58.5){$w$}
\put(0,50){\unitlength=1.2mm\line(1,0){61}}\put(62,49){1,0}
\put(0,50.2){\unitlength=1.2mm\line(1,0){60}}
\put(20,9.8){\unitlength=1.2mm\line(1,0){60}}
\put(60,30){\line(1,0){1}}\put(62,29){0,5} \end{picture}}

\vskip -5mm Fig. 3.

\vskip 30mm

\centerline{\unitlength=1.2mm\begin{picture}(87,57)
\put(0,10){\vector(1,0){87}}\put(78,5){$\varepsilon-\mu$}
\multiput(20,10.5)(0,2){20}{\line(0,1){1.4}}
\multiput(0,50.1)(2,0){30}{\line(1,0){1.4}}
\put(20,10){\line(0,-1){1}}\put(17,5){$-I$}
\put(40,10){\line(0,-1){1}}\put(36.9,5){$-\frac12\hh I$}
\put(60,10){\line(0,-1){1}}\put(59.2,5){0}
\put(60,10){\vector(0,1){50}}\put(62,58.5){$w$} 
\put(60,50){\line(1,0){1}} \put(62,49){1,0}
\put(60,30){\line(1,0){1}}\put(62,29){0,5}
\put(60,10){\unitlength=1.2mm\special{em:linewidth 0.3pt}
\put(11.62,0.3333)  {\special{em:moveto}} \put(9.528,0.6667)
{\special{em:lineto}} \put(8.160,1)  {\special{em:lineto}}
\put(5.360,2)  {\special{em:lineto}} \put(3.7,3)
{\special{em:lineto}} }
\put(60,10){\unitlength=1.2mm\special{em:linewidth 0.3pt}
\put(-43.5,37)  {\special{em:moveto}} \put(-43.5,37)
{\special{em:lineto}} \put(-45.36,38) {\special{em:lineto}}
\put(-48.16,39) {\special{em:lineto}} \put(-49.52,39.33)
{\special{em:lineto}} \put(-51.60,39.67) {\special{em:lineto}} }
\put(60,10){\unitlength=1.2mm\special{em:linewidth 0.3pt}
\put(-43.84,36.59)  {\special{em:moveto}} \put(-43.76,35.77)
{\special{em:lineto}} \put(-43.60,34.84)  {\special{em:lineto}}
\put(-43.40,33.82)  {\special{em:lineto}} \put(-43.16,32.73)
{\special{em:lineto}} \put(-42.88,31.57)  {\special{em:lineto}}
\put(-40.96,23.96)  {\special{em:lineto}} \put(-40.64,22.64)
{\special{em:lineto}} \put(-40.32,21.32)  {\special{em:lineto}}
\put(-39.98,19.99)  {\special{em:lineto}} \put(-39.66,18.66)
{\special{em:lineto}} \put(-39.32,17.33)  {\special{em:lineto}}
\put(-38.98,16.00)  {\special{em:lineto}} \put(-38.64,14.66)
{\special{em:lineto}} \put(-38.26,13.33)  {\special{em:lineto}}
\put(-37.88,12.00)  {\special{em:lineto}} \put(-37.47,10.66)
{\special{em:lineto}} \put(-37.02,9.336)  {\special{em:lineto}}
\put(-36.54,8.000)  {\special{em:lineto}} \put(-35.98,6.668)
{\special{em:lineto}} \put(-35.32,5.336)  {\special{em:lineto}}
\put(-34.51,4.000)  {\special{em:lineto}} \put(-33.41,2.664)
{\special{em:lineto}} \put(-32.06,1.600)  {\special{em:lineto}}
\put(-31.57,1.336)  {\special{em:lineto}} \put(-30.27,0.800)
{\special{em:lineto}} \put(-29.20,0.536)  {\special{em:lineto}}
\put(-27.56,0.264)  {\special{em:lineto}} \put(-1.100,0.004)
{\special{em:lineto}} \put(-0.344,0.004)  {\special{em:lineto}}
\put(0.3148,0.020)  {\special{em:lineto}} \put(0.9626,0.044)
{\special{em:lineto}} \put(1.6024,0.096)  {\special{em:lineto}}
\put(2.2572,0.208)  {\special{em:lineto}} \put(2.8704,0.428)
{\special{em:lineto}} \put(3.4012,0.845)  {\special{em:lineto}}
\put(3.7720,1.567)  {\special{em:lineto}} \put(3.8710,3.401)
{\special{em:lineto}} \put(3.7714,4.232)  {\special{em:lineto}}
\put(3.6118,5.162)  {\special{em:lineto}} \put(3.4014,6.181)
{\special{em:lineto}} \put(3.1508,7.275)  {\special{em:lineto}}
\put(2.8701,8.431)  {\special{em:lineto}} \put(2.5689,9.634)
{\special{em:lineto}} \put(2.2545,10.88)  {\special{em:lineto}}
\put(1.9330,12.14)  {\special{em:lineto}} \put(1.6077,13.43)
{\special{em:lineto}} \put(1.2823,14.73)  {\special{em:lineto}}
\put(0.9579,16.04)  {\special{em:lineto}} \put(0.6335,17.36)
{\special{em:lineto}} \put(0.3093,18.69)  {\special{em:lineto}}
\put(-0.014,20.01)  {\special{em:lineto}} \put(-0.344,21.34)
{\special{em:lineto}} \put(-0.669,22.67)  {\special{em:lineto}}
\put(-1.016,24.00)  {\special{em:lineto}} \put(-1.401,25.34)
{\special{em:lineto}} \put(-1.729,26.67)  {\special{em:lineto}}
\put(-2.127,28.00)  {\special{em:lineto}} \put(-2.523,29.33)
{\special{em:lineto}} \put(-2.899,30.67)  {\special{em:lineto}}
\put(-3.456,32.00)  {\special{em:lineto}} \put(-3.993,33.33)
{\special{em:lineto}} \put(-4.607,34.67)  {\special{em:lineto}}
\put(-5.484,36.00)  {\special{em:lineto}} \put(-6.555,37.33)
{\special{em:lineto}} \put(-7.088,37.87)  {\special{em:lineto}}
\put(-8.390,38.67)  {\special{em:lineto}} \put(-9.074,38.93)
{\special{em:lineto}} \put(-10.68,39.47)  {\special{em:lineto}}
\put(-12.56,39.73)  {\special{em:lineto}} \put(-38.98,40.00)
{\special{em:lineto}} \put(-39.66,40.00)  {\special{em:lineto}}
\put(-40.32,39.98)  {\special{em:lineto}} \put(-40.96,39.96)
{\special{em:lineto}} \put(-41.60,39.90)  {\special{em:lineto}}
\put(-42.24,39.79)  {\special{em:lineto}} \put(-42.88,39.57)
{\special{em:lineto}} \put(-43.40,39.16)  {\special{em:lineto}}
\put(-43.76,38.43)  {\special{em:lineto}} \put(-43.84,36.59)
{\special{em:lineto}} } \end{picture}}

\vskip -5mm Fig. 4.

$\phantom a$ \vskip 40mm

\centerline{\unitlength=1.2mm\begin{picture}(87,77)
\put(0,10){\vector(1,0){87}}\put(79,5){$\varepsilon-\mu$}
\multiput(20,10.5)(0,2){20}{\line(0,1){1.4}}
\multiput(0,50.5)(2,0){31}{\line(1,0){1.4}}
\put(20,10){\line(0,-1){1}}\put(17,5){$-I$}
\put(40,10){\line(0,-1){1}}\put(36.9,5){$-\frac12\hh I$}
\put(60,10){\line(0,-1){1}}\put(59.2,5){0}
\put(60,10){\vector(0,1){50}}\put(62,58.5){$w$} 
\put(62,49){1,0} \put(60,30){\line(1,0){1}}\put(62,29){0,5}
\put(60,10){\unitlength=1.2mm\special{em:linewidth 0.3pt}
\put(27.00,1.2) {\special{em:moveto}}
\put(25.99,1.3){\special{em:lineto}} \put(21.55,2)
{\special{em:lineto}} \put(17.10,3) {\special{em:lineto}}
\put(13.58,4)  {\special{em:lineto}} \put(10.57,5)
{\special{em:lineto}} \put(7.880,6) {\special{em:lineto}}
\put(5.408,7)  {\special{em:lineto}} \put(3.088,8)
{\special{em:lineto}} \put(0.8960,9) {\special{em:lineto}}
\put(-1.208,10)  {\special{em:lineto}} \put(-1.208,10)
{\special{em:lineto}} \put(-3.244,11) {\special{em:lineto}} }
\put(60,10){\unitlength=1.2mm\special{em:linewidth 0.3pt}
\put(-37,29) {\special{em:moveto}} \put(-36.76,29)
{\special{em:lineto}} \put(-38.79,30) {\special{em:lineto}}
\put(-40.88,31)  {\special{em:lineto}} \put(-43.08,32)
{\special{em:lineto}} \put(-45.40,33) {\special{em:lineto}}
\put(-47.88,34)  {\special{em:lineto}} \put(-50.56,35)
{\special{em:lineto}} \put(-53.56,36) {\special{em:lineto}}
\put(-57.08,37)  {\special{em:lineto}} \put(-60.00,37.7)
{\special{em:lineto}} }
\put(60,10){\unitlength=1.2mm\special{em:linewidth 0.3pt}
\put(-36.61,28.25)  {\special{em:moveto}} \put(-36.56,27.54)
{\special{em:lineto}} \put(-36.49,26.81)  {\special{em:lineto}}
\put(-36.30,26.00)  {\special{em:lineto}} \put(-36.16,25.22)
{\special{em:lineto}} \put(-35.90,24.36)  {\special{em:lineto}}
\put(-35.60,23.46)  {\special{em:lineto}} \put(-35.30,22.58)
{\special{em:lineto}} \put(-34.92,21.62)  {\special{em:lineto}}
\put(-34.49,20.64)  {\special{em:lineto}} \put(-34.04,19.66)
{\special{em:lineto}} \put(-33.52,18.62)  {\special{em:lineto}}
\put(-32.93,17.55)  {\special{em:lineto}} \put(-32.27,16.47)
{\special{em:lineto}} \put(-31.54,15.34)  {\special{em:lineto}}
\put(-30.70,14.16)  {\special{em:lineto}} \put(-29.75,12.97)
{\special{em:lineto}} \put(-28.64,11.70)  {\special{em:lineto}}
\put(-27.28,10.36)  {\special{em:lineto}} \put(-25.57,8.904)
{\special{em:lineto}} \put(-23.82,7.696)  {\special{em:lineto}}
\put(-22.74,7.064)  {\special{em:lineto}} \put(-20.56,6.016)
{\special{em:lineto}} \put(-17.259,4.936)  {\special{em:lineto}}
\put(-13.507,4.348)  {\special{em:lineto}} \put(-11.362,4.300)
{\special{em:lineto}} \put(-9.7588,4.428)  {\special{em:lineto}}
\put(-8.4684,4.664)  {\special{em:lineto}} \put(-7.3924,4.988)
{\special{em:lineto}} \put(-6.4840,5.376)  {\special{em:lineto}}
\put(-5.7180,5.848)  {\special{em:lineto}} \put(-5.0736,6.376)
{\special{em:lineto}} \put(-4.5384,6.984)  {\special{em:lineto}}
\put(-4.1092,7.640)  {\special{em:lineto}} \put(-3.7788,8.388)
{\special{em:lineto}} \put(-3.5472,9.188)  {\special{em:lineto}}
\put(-3.4080,10.12)  {\special{em:lineto}} \put(-3.379, 11.74)
{\special{em:lineto}} \put(-3.4399,12.46)  {\special{em:lineto}}
\put(-3.5434,13.20)  {\special{em:lineto}} \put(-3.6889,13.99)
{\special{em:lineto}} \put(-3.8770,14.80)  {\special{em:lineto}}
\put(-4.1085,15.65)  {\special{em:lineto}} \put(-4.3842,16.53)
{\special{em:lineto}} \put(-4.7046,17.43)  {\special{em:lineto}}
\put(-5.0727,18.38)  {\special{em:lineto}} \put(-5.4901,19.35)
{\special{em:lineto}} \put(-5.9590,20.35)  {\special{em:lineto}}
\put(-6.4850,21.38)  {\special{em:lineto}} \put(-7.0731,22.44)
{\special{em:lineto}} \put(-7.7300,23.54)  {\special{em:lineto}}
\put(-8.4674,24.67)  {\special{em:lineto}} \put(-9.3005,25.83)
{\special{em:lineto}} \put(-10.252,27.04)  {\special{em:lineto}}
\put(-11.364,28.30)  {\special{em:lineto}} \put(-12.705,29.63)
{\special{em:lineto}} \put(-14.444,31.10)  {\special{em:lineto}}
\put(-15.347,31.76)  {\special{em:lineto}} \put(-16.176,32.30)
{\special{em:lineto}} \put(-17.259,32.94)  {\special{em:lineto}}
\put(-19.453,33.99)  {\special{em:lineto}} \put(-22.74,35.06)
{\special{em:lineto}} \put(-26.49,35.65)  {\special{em:lineto}}
\put(-28.64,35.70)  {\special{em:lineto}} \put(-30.24,35.57)
{\special{em:lineto}} \put(-31.54,35.34)  {\special{em:lineto}}
\put(-32.61,35.01)  {\special{em:lineto}} \put(-33.52,34.62)
{\special{em:lineto}} \put(-34.28,34.15)  {\special{em:lineto}}
\put(-34.92,33.62)  {\special{em:lineto}} \put(-35.45,33.02)
{\special{em:lineto}} \put(-35.90,32.36)  {\special{em:lineto}}
\put(-36.23,31.61)  {\special{em:lineto}} \put(-36.49,30.81)
{\special{em:lineto}} \put(-36.55,29.88)  {\special{em:lineto}}
\put(-36.61,28.25)  {\special{em:lineto}} } \end{picture}}

\vskip -5mm Fig. 5.

\vskip 30mm

\unitlength=1.2mm \centerline{\begin{picture}(79,57)
\put(12,10){\vector(1,0){59.5}}\put(69.5,5){$\tau$}
\put(12,10){\line(0,-1){1}}\put(11.2,5){0}
\put(32,10){\line(0,-1){1}}\put(29.7,5){0,5}
\put(52,10){\line(0,-1){1}}\put(51.2,5){1}
\put(12,10){\vector(0,1){50}}\put(14,57.4){$d_{\hh max} (\tau)$}
\put(12,10){\line(-1,0){1}}\put(4,9){$0$}
\put(12,30){\line(-1,0){1}}\put(4,29){$0,5$}
\put(12,50){\line(-1,0){1}}\put(4,49){$1,0$}
\put(12,10){\unitlength=1.2mm\special{em:linewidth 0.3pt}
\put(40.000,0)  {\special{em:moveto}} \put(39.967,2)
{\special{em:lineto}} \put(39.886,4)  {\special{em:lineto}}
\put(39.698,6)  {\special{em:lineto}} \put(39.461,8)
{\special{em:lineto}} \put(39.152,10)  {\special{em:lineto}}
\put(38.770,12)  {\special{em:lineto}} \put(38.310,14)
{\special{em:lineto}} \put(37.767,16)  {\special{em:lineto}}
\put(37.136,18)  {\special{em:lineto}} \put(36.410,20)
{\special{em:lineto}} \put(35.577,22)  {\special{em:lineto}}
\put(34.625,24)  {\special{em:lineto}} \put(33.536,26)
{\special{em:lineto}} \put(32.284,28)  {\special{em:lineto}}
\put(30.834,30)  {\special{em:lineto}} \put(29.128,32)
{\special{em:lineto}} \put(27.067,34)  {\special{em:lineto}}
\put(24.453,36)  {\special{em:lineto}} \put(23.830,36.40)
{\special{em:lineto}} \put(23.159,36.80)  {\special{em:lineto}}
\put(22.431,37.20)  {\special{em:lineto}} \put(21.634,37.60)
{\special{em:lineto}} \put(20.745,38)  {\special{em:lineto}}
\put(19.734,38.40)  {\special{em:lineto}} \put(18.544,38.80)
{\special{em:lineto}} \put(17.062,39.20)  {\special{em:lineto}}
\put(16.76,39.20)  {\special{em:lineto}} \put(14.80,39.60)
{\special{em:lineto}} \put(13.20,39.80)  {\special{em:lineto}}
\put(10.50,39.96)  {\special{em:lineto}} \put(0,40)
{\special{em:lineto}} } \end{picture}}

\vskip -5mm Fig. 6.

\vskip 20mm

\unitlength=1.2mm\centerline{\begin{picture}(85,80)\put(23,2){
1}\put(27,11){\it 2}\put(30,19){\it 3}\put(7,25){\it 1}
\put(0,35){\vector(1,0){85}}\put(78,31.7){$\varepsilon$\hs--\hs$\mu$}
\put(61,32){0} \multiput(20,-5)(0,2.1){20}{\line(0,1){1.4}}
\put(20,35){\line(0,1){1}}\put(14,36){$-\hh I$}
\put(40,35){\line(0,1){1}}\put(38,38){$-\hh\frac{I}{\hh 2\hh}$}
\put(60,15){\line(1,0){1}}\put(62,14){$-\hh\frac{I}{\hh 2\hh}$}
\put(60,-10){\vector(0,1){71}}\put(62,58){$\overline\varepsilon$\hs--\hs$\mu$}
\multiput(20,35)(1.4,1.4){18}{\unitlength=1.2mm\special{em:linewidth
0.3pt} \put(0,0){\special{em:moveto}}\put(1,1){\special{em:lineto}}
}\put(0,15){\line(1,1){20}}\put(20,-5){\line(1,1){61}}
\multiput(20,-5)(2.1,0){20}{\line(1,0){1.4}}\put(62,-6){$-\hh{I}$}
\put(60,35){\unitlength=1.2mm\special{em:linewidth 0.3pt}
\put(21.000,23.20){\special{em:moveto}}
\put(19.165,21.66){\special{em:lineto}}
\put(17.098,20.10){\special{em:lineto}}
\put(15.257,18.76){\special{em:lineto}}
\put(13.578,17.58){\special{em:lineto}}
\put(12.024,16.53){\special{em:lineto}}
\put(10.567,15.57){\special{em:lineto}}
\put(9.1900,14.69){\special{em:lineto}}
\put(7.8770,13.88){\special{em:lineto}}
\put(6.6180,13.12){\special{em:lineto}}
\put(5.4050,12.41){\special{em:lineto}}
\put(4.2300,11.73){\special{em:lineto}}
\put(3.0900,11.09){\special{em:lineto}}
\put(1.9790,10.48){\special{em:lineto}}
\put(0.8936,9.896){\special{em:lineto}}
\put(-0.169,9.328){\special{em:lineto}}
\put(-1.211,8.792){\special{em:lineto}}
\put(-2.236,8.264){\special{em:lineto}}
\put(-3.245,7.755){\special{em:lineto}} }
\put(60,35){\unitlength=1.2mm\special{em:linewidth 0.3pt}
\put(-36.755,-7.755){\special{em:moveto}}
\put(-37.764,-8.264){\special{em:lineto}}
\put(-38.789,-8.792){\special{em:lineto}}
\put(-39.831,-9.328){\special{em:lineto}}
\put(-40.894,-9.896){\special{em:lineto}}
\put(-41.979,-10.48){\special{em:lineto}}
\put(-43.090,-11.09){\special{em:lineto}}
\put(-44.230,-11.73){\special{em:lineto}}
\put(-45.405,-12.41){\special{em:lineto}}
\put(-46.618,-13.12){\special{em:lineto}}
\put(-47.877,-13.88){\special{em:lineto}}
\put(-49.190,-14.69){\special{em:lineto}}
\put(-50.567,-15.57){\special{em:lineto}}
\put(-52.024,-16.53){\special{em:lineto}}
\put(-53.578,-17.58){\special{em:lineto}}
\put(-55.257,-18.76){\special{em:lineto}}
\put(-57.098,-20.10){\special{em:lineto}}
\put(-60.000,-22.20){\special{em:lineto}} }
\put(60,35){\unitlength=1.2mm\special{em:linewidth 0.3pt}
\put(-36.644,-7.700){\special{em:moveto}}
\put(-36.592,-8.668){\special{em:lineto}}
\put(-36.538,-9.164){\special{em:lineto}}
\put(-36.454,-9.680){\special{em:lineto}}
\put(-36.350,-10.14){\special{em:lineto}}
\put(-36.222,-10.62){\special{em:lineto}}
\put(-36.072,-11.08){\special{em:lineto}}
\put(-35.892,-11.54){\special{em:lineto}}
\put(-35.691,-11.99){\special{em:lineto}}
\put(-35.461,-12.43){\special{em:lineto}}
\put(-35.209,-12.88){\special{em:lineto}}
\put(-34.927,-13.30){\special{em:lineto}}
\put(-34.620,-13.72){\special{em:lineto}}
\put(-34.282,-14.12){\special{em:lineto}}
\put(-33.915,-14.52){\special{em:lineto}}
\put(-33.515,-14.90){\special{em:lineto}}
\put(-33.080,-15.25){\special{em:lineto}}
\put(-32.608,-15.60){\special{em:lineto}}
\put(-32.094,-15.91){\special{em:lineto}}
\put(-31.532,-16.20){\special{em:lineto}}
\put(-30.920,-16.45){\special{em:lineto}}
\put(-30.241,-16.67){\special{em:lineto}}
\put(-29.486,-16.84){\special{em:lineto}}
\put(-28.637,-16.93){\special{em:lineto}}
\put(-27.658,-16.94){\special{em:lineto}}
\put(-26.492,-16.84){\special{em:lineto}}
\put(-25.951,-16.74){\special{em:lineto}}
\put(-25.013,-16.52){\special{em:lineto}}
\put(-23.824,-16.13){\special{em:lineto}}
\put(-22.742,-15.68){\special{em:lineto}}
\put(-21.980,-15.31){\special{em:lineto}}
\put(-20.548,-14.55){\special{em:lineto}}
\put(-19.453,-13.85){\special{em:lineto}}
\put(-18.020,-12.88){\special{em:lineto}}
\put(-17.259,-12.31){\special{em:lineto}}
\put(-16.176,-11.47){\special{em:lineto}}
\put(-14.986,-10.49){\special{em:lineto}}
\put(-14.048,-9.660){\special{em:lineto}}
\put(-13.507,-9.168){\special{em:lineto}}
\put(-12.341,-8.060){\special{em:lineto}}
\put(-11.362,-7.060){\special{em:lineto}}
\put(-10.513,-6.168){\special{em:lineto}}
\put(-9.7588,-5.340){\special{em:lineto}}
\put(-9.0808,-4.548){\special{em:lineto}}
\put(-8.4684,-3.800){\special{em:lineto}}
\put(-7.9056,-3.100){\special{em:lineto}}
\put(-7.3924,-2.408){\special{em:lineto}}
\put(-6.9196,-1.748){\special{em:lineto}}
\put(-6.4840,-1.100){\special{em:lineto}}
\put(-6.0848,-0.488){\special{em:lineto}}
\put(-5.7180,0.1200){\special{em:lineto}}
\put(-5.3788,0.7200){\special{em:lineto}}
\put(-5.0736,1.3000){\special{em:lineto}}
\put(-4.7916,1.8720){\special{em:lineto}}
\put(-4.5384,2.4200){\special{em:lineto}}
\put(-4.3088,2.9920){\special{em:lineto}}
\put(-4.1092,3.5400){\special{em:lineto}}
\put(-3.9280,4.0720){\special{em:lineto}}
\put(-3.7788,4.6200){\special{em:lineto}}
\put(-3.6508,5.1400){\special{em:lineto}}
\put(-3.5472,5.6800){\special{em:lineto}}
\put(-3.4620,6.1600){\special{em:lineto}}
\put(-3.4080,6.6640){\special{em:lineto}}
\put(-3.3556,7.1040){\special{em:lineto}}
\put(-3.3556,7.7000){\special{em:lineto}} }
\put(60,35){\unitlength=1.2mm\special{em:linewidth 0.3pt}
\put(21.347,21.85){\special{em:moveto}}
\put(17.318,18.32){\special{em:lineto}}
\put(14.726,16.23){\special{em:lineto}}
\put(12.722,14.72){\special{em:lineto}}
\put(11.041,13.54){\special{em:lineto}}
\put(9.5620,12.56){\special{em:lineto}}
\put(8.2230,11.73){\special{em:lineto}}
\put(6.9860,10.99){\special{em:lineto}}
\put(5.8280,10.33){\special{em:lineto}}
\put(4.7295,9.730){\special{em:lineto}}
\put(3.6810,9.180){\special{em:lineto}}
\put(2.7730,8.675){\special{em:lineto}} }
\put(60,35){\unitlength=1.2mm\special{em:linewidth 0.3pt}
\put(-42.900,-8.675){\special{em:moveto}}
\put(-43.681,-9.180){\special{em:lineto}}
\put(-44.730,-9.730){\special{em:lineto}}
\put(-45.828,-10.33){\special{em:lineto}}
\put(-46.986,-10.99){\special{em:lineto}}
\put(-48.223,-11.73){\special{em:lineto}}
\put(-49.562,-12.56){\special{em:lineto}}
\put(-51.041,-13.54){\special{em:lineto}}
\put(-52.722,-14.72){\special{em:lineto}}
\put(-54.726,-16.23){\special{em:lineto}}
\put(-57.318,-18.32){\special{em:lineto}}
\put(-60.000,-20.55){\special{em:lineto}} }
\put(60,35){\unitlength=1.2mm\special{em:linewidth 0.3pt}
\put(-42.904,-8.8000){\special{em:moveto}}
\put(-42.904,-9.8224){\special{em:lineto}}
\put(-42.768,-10.859){\special{em:lineto}}
\put(-42.540,-11.918){\special{em:lineto}}
\put(-42.224,-12.993){\special{em:lineto}}
\put(-41.828,-14.084){\special{em:lineto}}
\put(-41.360,-15.190){\special{em:lineto}}
\put(-40.820,-16.299){\special{em:lineto}}
\put(-40.220,-17.412){\special{em:lineto}}
\put(-39.558,-18.520){\special{em:lineto}}
\put(-38.836,-19.612){\special{em:lineto}}
\put(-38.048,-20.678){\special{em:lineto}}
\put(-37.190,-21.704){\special{em:lineto}}
\put(-36.245,-22.670){\special{em:lineto}}
\put(-35.192,-23.549){\special{em:lineto}}
\put(-33.992,-24.300){\special{em:lineto}}
\put(-32.574,-24.850){\special{em:lineto}}
\put(-31.566,-25.040){\special{em:lineto}}
\put(-30.799,-25.064){\special{em:lineto}}
\put(-29.434,-24.890){\special{em:lineto}}
\put(-28.307,-24.597){\special{em:lineto}}
\put(-27.644,-24.350){\special{em:lineto}}
\put(-25.976,-23.53){\special{em:lineto}}
\put(-24.841,-22.870){\special{em:lineto}}
\put(-22.512,-21.230){\special{em:lineto}}
\put(-20.850,-19.860){\special{em:lineto}}
\put(-19.153,-18.42){\special{em:lineto}}
\put(-16.776,-16.270){\special{em:lineto}}
\put(-15.159,-14.750){\special{em:lineto}}
\put(-14.022,-13.660){\special{em:lineto}}
\put(-12.358,-12.070){\special{em:lineto}}
\put(-11.102,-10.820){\special{em:lineto}}
\put(-9.2012,-8.9368){\special{em:lineto}}
\put(-7.4260,-7.1500){\special{em:lineto}}
\put(-6.0064,-5.6988){\special{em:lineto}}
\put(-4.8068,-4.4500){\special{em:lineto}}
\put(-3.7528,-3.3280){\special{em:lineto}}
\put(-2.8080,-2.2940){\special{em:lineto}}
\put(-1.9504,-1.3208){\special{em:lineto}}
\put(-1.1644,-0.3880){\special{em:lineto}}
\put(-0.4400,0.52120){\special{em:lineto}}
\put(0.22160,1.41400){\special{em:lineto}}
\put(0.82120,2.30000){\special{em:lineto}}
\put(1.35960,3.18920){\special{em:lineto}}
\put(1.82880,4.08520){\special{em:lineto}}
\put(2.22400,4.99320){\special{em:lineto}}
\put(2.53920,5.91720){\special{em:lineto}}
\put(2.77000,6.86120){\special{em:lineto}}
\put(2.90520,8.82360){\special{em:lineto}} } \end{picture}}

\vskip 20mm Fig. 7.

\newpage

\hskip -13pt{\Large\bf The Fermi -- Dirac function and }

\vskip 1mm \hskip -13pt{\Large\bf the anisotropic distribution
 of the interacting electrons}

\vskip 3 mm \hskip -13pt{\large Boris V. Bondarev }

\vskip 5mm \hskip -13pt

\hskip -13pt Fig. 1. The isotropic distribution function of the
conduction electrons over the energy at the different values of
the temperature $\tau$: 1 -- $\tau=0$; 2 -- $\tau=0.25$; 3 --
$\tau=0.8$.

\vskip 5mm \hskip -13pt

\hskip -13pt Fig. 2. The anisotropic distribution function of the
conduction electrons over the energy at the different values of
the temperature $\tau$: 1 -- $\tau=0$; 2 -- $\tau=0.25$; 3 --
$\tau=0.8$; 4 -- $\tau=0.95$.

\vskip 5mm \hskip -13pt

\hskip -13pt Fig. 3 . The anisotropic distribution function of the
conduction electrons over the energy at the different values of the
temperature $\tau=0$ with the lowest energy.

\vskip 5mm \hskip -13pt

\hskip -13pt Fig. 4 . The anisotropic distribution function of the
conduction electrons over the energy at the different values of the
temperature $\tau=0.25$ with the lowest energy.

\vskip 5mm \hskip -13pt

\hskip -13pt Fig. 5 . The anisotropic distribution function of the
conduction electrons over the energy at the different values of the
temperature $\tau=0.8$ with the lowest energy.

\vskip 5mm \hskip -13pt

\hskip -13pt Fig. 6. The anisotropy parameter of the electron
distribution $d_{\hh max}$ as a function of the temperature $\tau$.

\vskip 5mm \hskip -13pt

\hskip -13pt Fig. 7. The dependence of the average electron energy
$\overline\varepsilon$ on its kinetic energy $\varepsilon$ for the
temperature $\tau$: 1 -- $\tau=0$; 2 -- $\tau=0.5$; 3 --
$\tau=0.8$.

\end{document}